\documentstyle[proceedings,epsfig]{crckapb}
\def\deg{{$^{\circ}$}}
\begin{opening}
\title
{Orientation effects on bent extragalactic jets}

\author{Stephen Higgins} 
\institute{Manchester Metropolitan University, Chester Street, Manchester, UK}

\author{Tim O'Brien}
\institute
{University of Manchester, Jodrell Bank, Macclesfield, UK}

\author{James Dunlop}
\institute
{Institute for Astronomy, University of Edinburgh, Blackford Hill, Edinburgh}
\runningauthor{S.\ Higgins, T.\ O'Brien, J.\ Dunlop}
\runningtitle{Orientation effects on bent extragalactic jets}

\end{opening}

\begin{document}
\parskip=2mm
\baselineskip=5.5mm

\setlength{\unitlength}{1.0cm}

\begin{abstract}
We have investigated how varying several parameters affects the
results of a collision between an extragalactic jet and a dense,
intergalactic cloud, through a series of hydrodynamic simulations.  We
have produced synthetic radio images for comparison with observations.
These show that a variety of structures may be produced from simple
jet-cloud collisions. Moderate Mach numbers and density contrasts are
needed to produce observable bends. We investigate the effect of
viewing from various angles on the appearance of such sources.

\end{abstract}

\section
{Observations of distorted jets}

The jets and hotspots of radio galaxies and quasars often show complex
structure.  Jets
can bend by over 90$^{\circ}$ and remain collimated for several jet
radii \cite{BP84}, despite the expectation that the
oblique shock causing the bend should decelerate the jet \cite{I91}. 
Barthel {\it et al.} \cite{BMSL88} present a large sample of quasars 
in which 25\% showed bending greater than 20\deg. 
Explanations for these complex structures include collision with dense
clouds in the ambient medium \cite{SBC85,LB98}.

In studies of astrophysical fluid dynamics  we
can only measure the emission properties of sources as projected onto
the sky. We must infer flow properties from such observations. In
 numerical simulations we must attempt to
invert this by estimating the emission properties of our solutions and
considering the effect of the projection of three-dimensional
solutions onto two-dimensional observations. We present the results of
such studies, along with some tentative conclusions, in section \ref{Results}.

\section{Jets and their interaction with environment}

\label{numsims}

Much of the understanding of the fluid dynamics of extragalactic jets,
and of the shape and structures found in observations, has come from
numerical simulations \cite{W91}. Fully three-dimensional
simulations have only become possible fairly recently \cite{N93},
but many important results have been obtained from axisymmetric calculations. 
Several models for the production of complex and distorted structures
in radio jets and lobes have been explored through numerical
simulations including: variations in the direction of the jet at its
source \cite{WG85,S82}; cross-winds \cite{leahy84}; 
the source axis is not parallel to the axis of
a spheroidal gas distribution, or the source galaxy moves through the
cluster medium \cite{LW84}; helical
instabilities \cite{S97}; oblique magnetic fields in the intra-cluster
medium \cite{K96}. Cloud collisions are particularly applicable in
cases where these bends are very sharp. Loken {\it et al.} \cite{L95} show
that the necessary gas velocities 
  can arise in cluster mergers, as can shocks that will bend the jet.
These simulations still do not explain the sharpness of the bend.


The first investigation of the effect of off-axis jet-cloud collisions
was by De Young \cite{DY91} using the `beam scheme' \cite{SP74}. 
De Young 
observed that the jet was decelerated by the cloud. 
 A similar interaction was investigated at higher resolution by Balsara and
Norman using their RIEMANN code \cite{N93}. They argued 
 that a De Laval nozzle was formed which 
re-accelerated the jet in a new direction after impact. They did not present
any results at later time to show the formation of a deflected flow pattern.

More recently Raga and Canto \cite{RC96} have published analytical
calculations and two-dimensional simulations showing bending by
clouds. They conclude that slower jets will be bent more, and clouds
will be eroded as jets bore through them.

\section{Numerical methods}


We have extended this work 
 through a series of simulations using various sets of 
parameters \cite{H98,HOD99}. 
The parameters are: Mach number, and the 
density contrasts between of the jet and the cloud with the ambient
medium. Details of the simulations are given in table \ref{pstbl}.
We have assumed conditions in the ambient medium
consistent with observations: a temperature of 5$\times 10^7$K and a
particle number density of 0.01 cm$^{-3}$. These values are used to
form dimensionless units in the computation so that model values for
the ambient density and pressure in the code are set to 1.0. The jet
and cloud are both taken to be in pressure balance with the ambient
medium. 
\begin{table}[!b]
\caption{The values of the parameters for different simulations. 
There is no cloud in simulations 9 -- 12.  }
\label{pstbl}
\begin{tabular}{lllll}\hline
Simulation & Jet density & Cloud density & Jet mach & Jet \\
 number & contrast & contrast & number & speed\\
\hline
1, 2, 9 & 0.01  & 50,  200, -- & 2  & 0.07c \\
3, 4, 10 & 0.01  & 50, 200, -- & 10 & 0.36c \\
5, 6, 11 & 0.2   & 50, 200, -- & 2  & 0.02c \\
7, 8, 12 & 0.2   & 50, 200, --  & 10 & 0.08c \\
\hline
\end{tabular}
\end{table}


\label{synchro}

To calculate the synchrotron emission we need to express the
magnetic field and the energy distribution in terms of the results of
our hydrodynamic simulations. 
 We can then produce
synthetic radio maps by integrating this through the grid. We used the
data visualization package PV-Wave to examine the simulations. This has
the facility to rotate three-dimensional data sets, and hence integrate
along any chosen line of sight.

\section{Results}

\label{Results}

The interactions produce a variety of structures depending on the
values of these parameters, so this model can be applied to many radio
structures. Different structures can also be produced by a single set
of parameters as the interaction progresses. 
 Strong deflections ($\sim$~90\deg) are difficult to sustain, producing
transient structures with complex features such as double
hotspots. Deflection may be easier to produce or detect in lower power
jets close to the plane of the sky. This is the case for simulations 3 and 4.
As the jet breaks through the cloud there are two hotspots within a
boot-shaped lobe. This is a similar radio structure to 4C~29.50 \cite{4c2950}, 
with the close double hotspot. Although the cloud
impact is the cause of the bending, the deflection and secondary
hotspot is actually produced as the jet bends inside the distorted
cocoon that has been formed during the interaction.

It is difficult to reach firm conclusions on the basis of these
simulations of the kind of sources,
and how many of any kind, we would expect to observe. We need to know
how sources are distributed over the ranges of parameters, how the
environments vary in clumpiness and density and hence what the
probability of colliding with a density enhancement is.
However we can make crude estimates of the likely distribution of
sources by assuming fairly uniform distribution of the parameters
characterising the jets and their environments.
Figure~\ref{jet4cont}~shows contour plots of the radio intensity of
simulation~4 at two epochs (at $t=4$~and~8). These show that the
secondary hotspot that forms after impact is about one order of
magnitude fainter than the primary hotspot at the impact.  The line
connecting the two hotspots is about 90\deg~to the axis of the initial
jet direction, which would be interpreted in observations as a
90\deg~bend. It is about one jet radius long, but the observed width of
the jet is smaller than the real width, due to limb darkening, so in
observations this might be interpreted as a few jet radii. By the next
epoch the secondary hotspot has faded by an order of magnitude or so,
while the angle has increased to 120\deg.
\begin{figure}[!h]
\centerline{\hspace{-13cm}
\psfig{figure=jet4040_r_cont.eps, angle=90, height=60mm}
\psfig{figure=jet4080_r_cont.eps, angle=90, height=60mm}}
\caption[Radio intensity of simulation~4]{Contour plots of the radio 
intensity of simulation~4 at $t=4$~and~8. \label{jet4cont}}
\end{figure}

Clearly this would be difficult to detect without huge dynamic range
(signal to noise), and the 90\deg~structure only lasts for at most a
single epoch of the interaction. This is no more than 15\% of the
lifetime of the interaction. This interaction is itself short lived,
perhaps 10\% of the typical lifetime ofa radio source (10$^8$~years),
so we would expect only 1--2\% of sources with these parameters, 
{\em viewed in the plane of the sky}, to show 90\deg~bends.

\begin{figure}[!hb]
\centerline{\psfig{figure=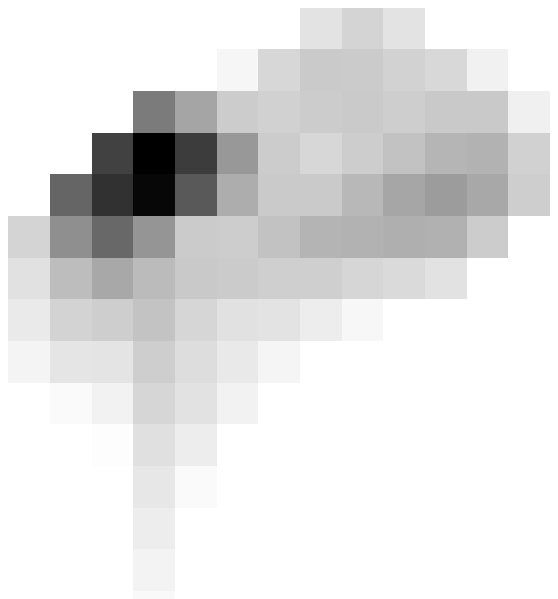, width=33mm}
\psfig{figure=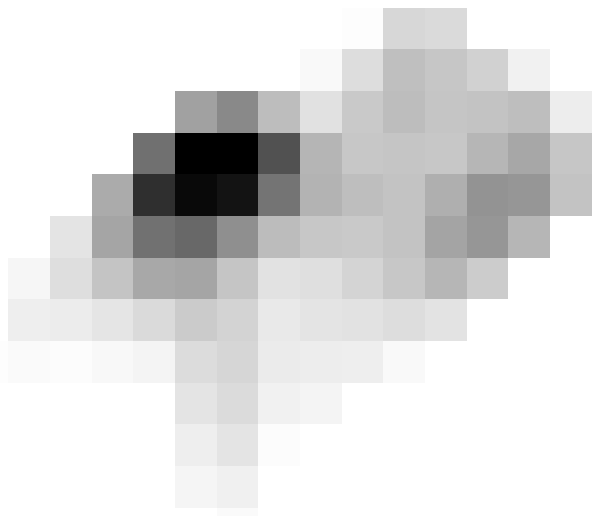, width=33mm}
\psfig{figure=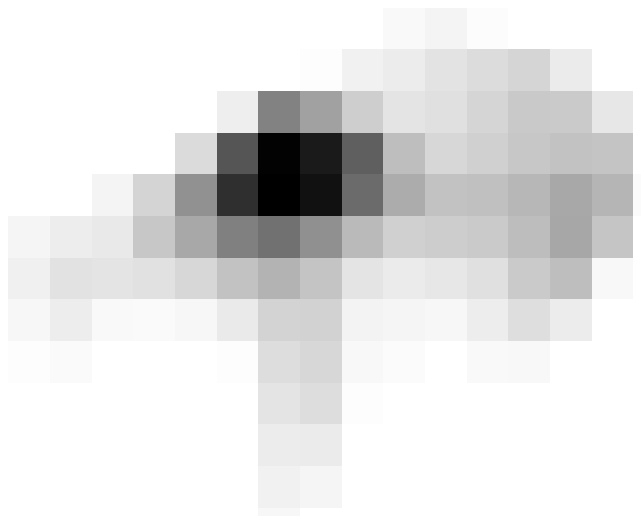, width=33mm}
\psfig{figure=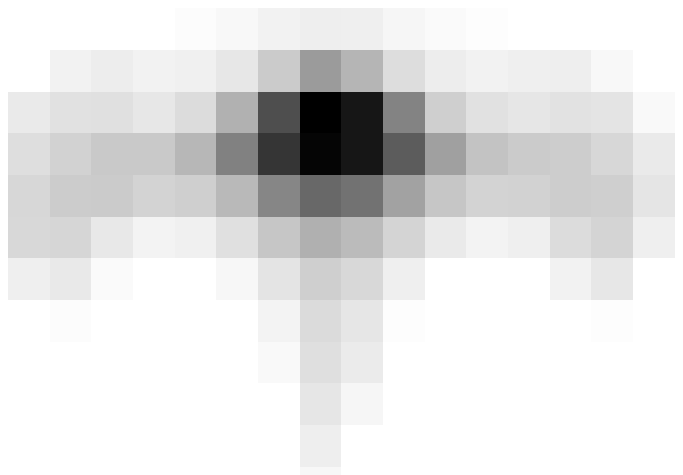, width=33mm}}

\vspace{-7mm}

\centerline{\psfig{figure=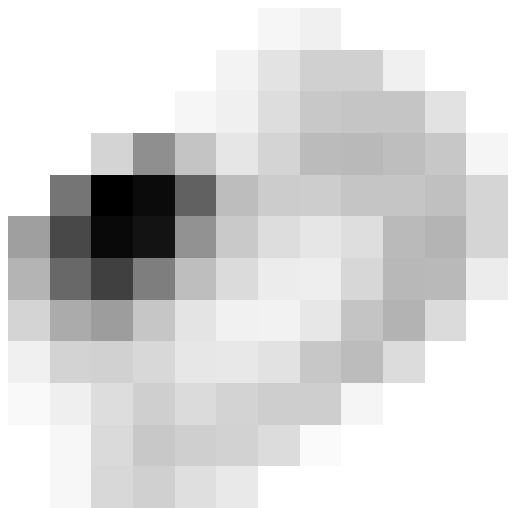, width=33mm}
\psfig{figure=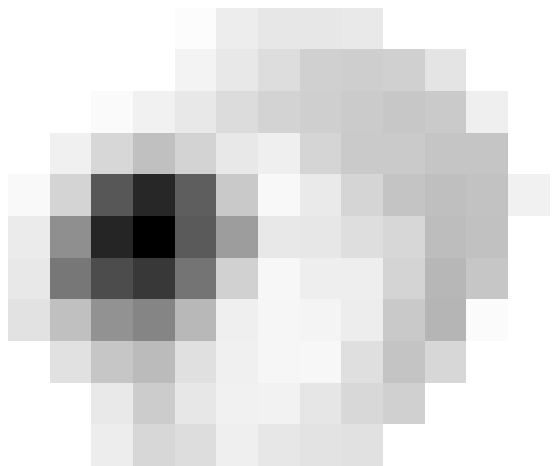, width=33mm}
\psfig{figure=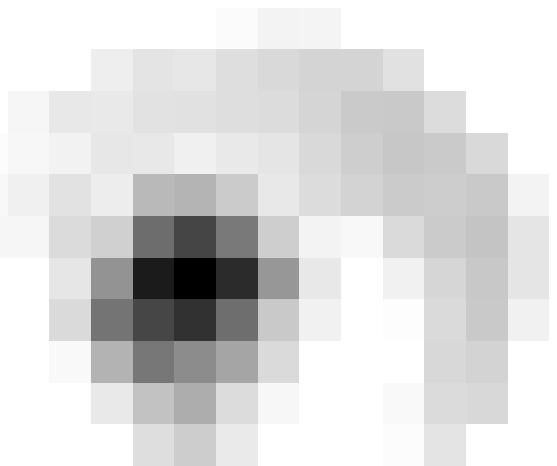, width=33mm}
\psfig{figure=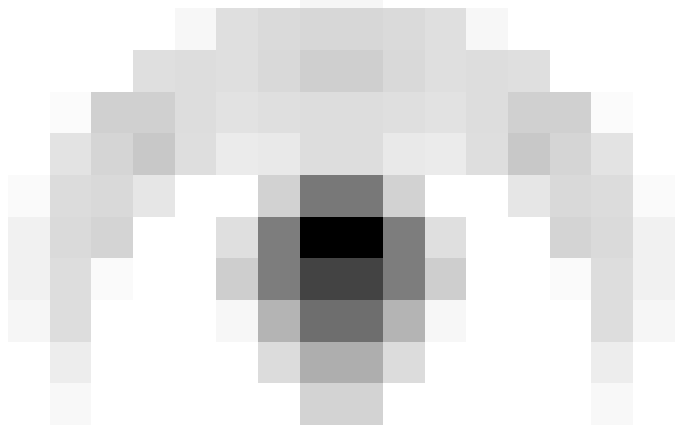, width=33mm}}

\vspace{3mm}

\centerline{\psfig{figure=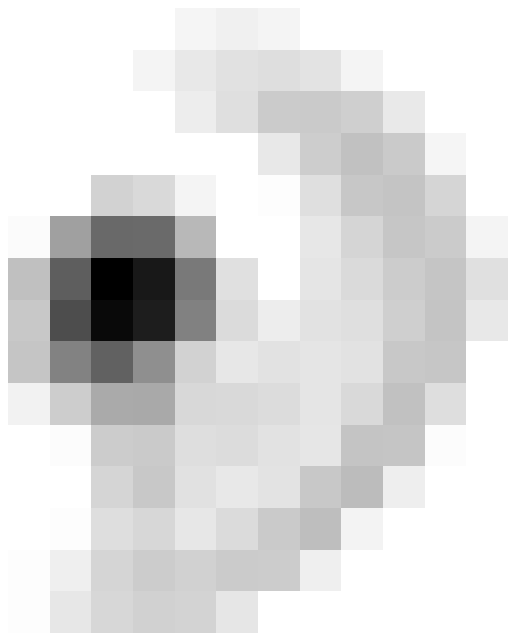, width=33mm}
\psfig{figure=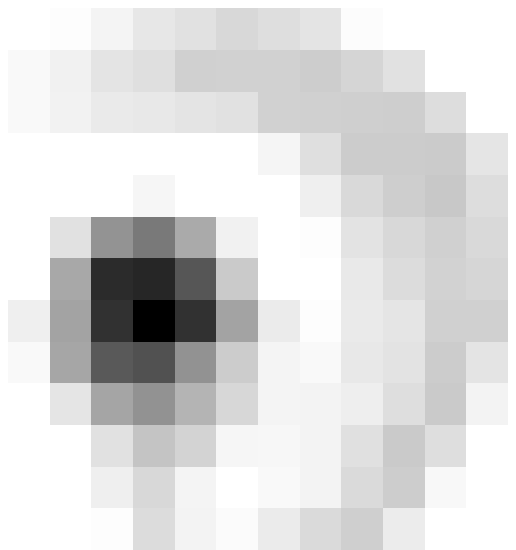, width=33mm}
\psfig{figure=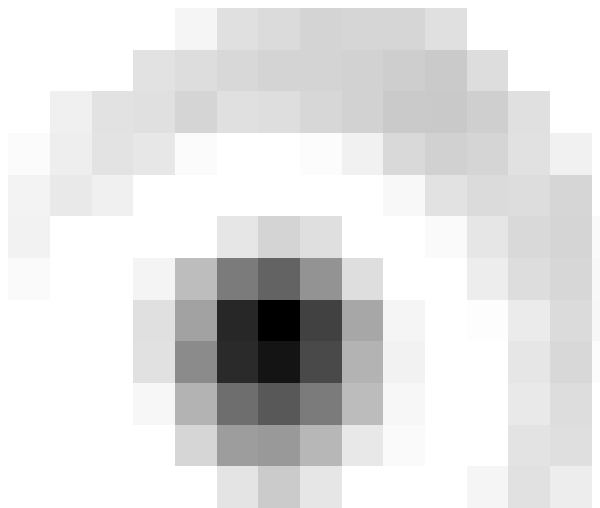, width=33mm}
\psfig{figure=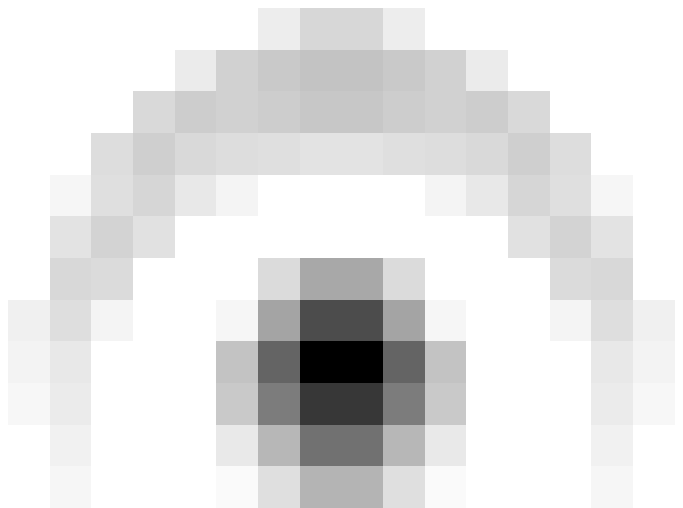, width=33mm}}

\vspace{3mm}

\centerline{\psfig{figure=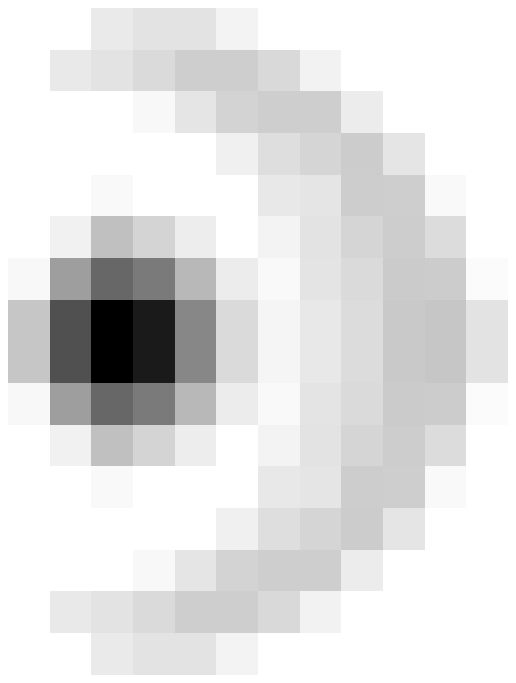, width=33mm}
\psfig{figure=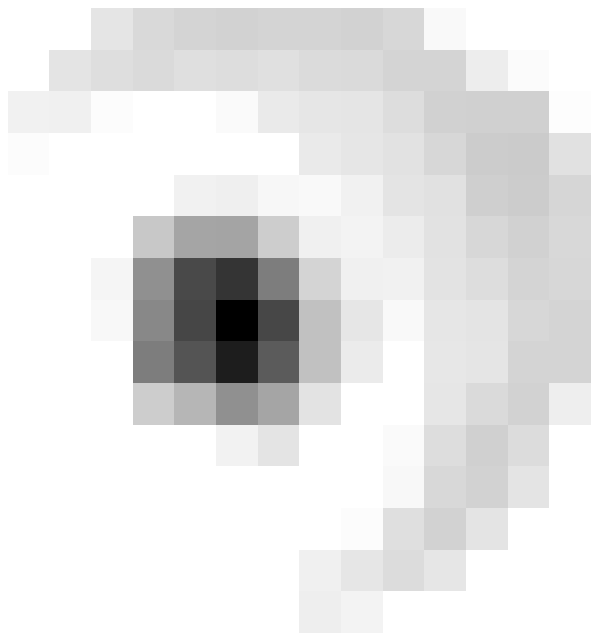,  width=33mm}
\psfig{figure=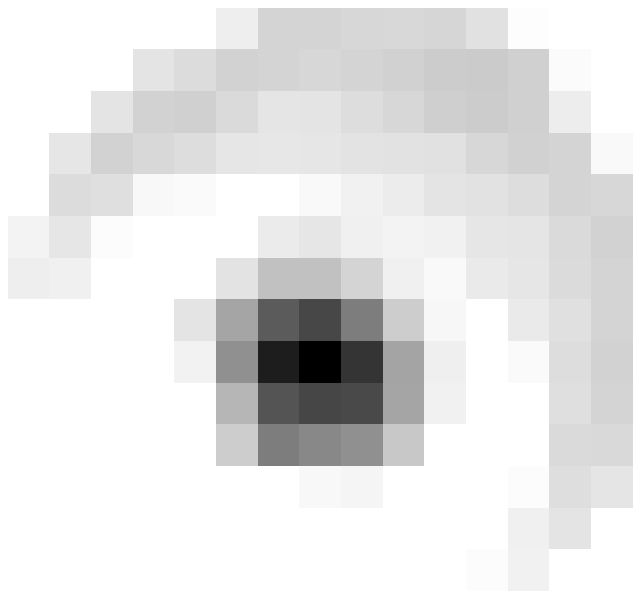, width=33mm}
\psfig{figure=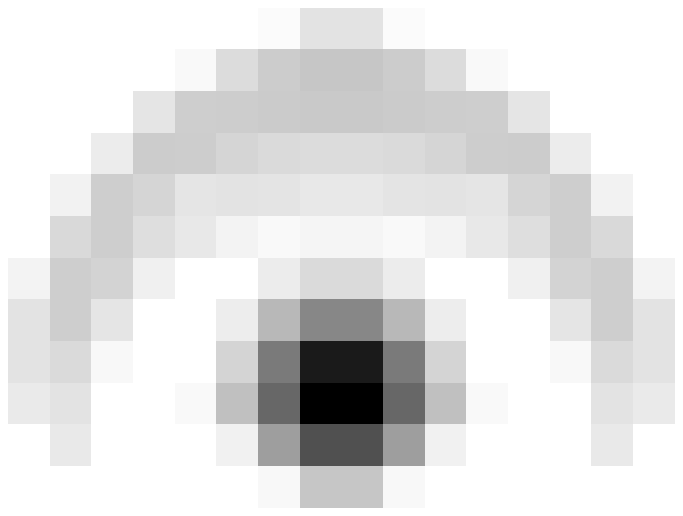, width=33mm}}

\vspace{3mm}

\caption[Integrated radio emission from simulation 4 at various angles.]
{ Integrated radio emission from simulation 4 
($M=2,\eta_j=0.01,\eta_c=200.0$) at various angles.}
\label{jet4rot}
\end{figure}

Figure~\ref{jet4rot}~shows the radio emission from simulation~4 at $t=4$
at several orientations.  Each column shows the source rotated by
30\deg~intervals, and each row is tilted 30\deg~toward the line of
sight.  The 90\deg~bend is only visible for a few
orientations. Assuming such sources are distributed isotropically with
viewing angle, we would only expect to detect about~20\%. Thus we
would only expect to see between a third and a half~a~percent of sources
with these parameters.

These simulations are the only one of the four sets of jet parameters
that show a 90\deg~deflection with a secondary hotspot. If this is
a representative sampling of jet parameters then we would
expect at most a quarter of all sources to fall into this region. Thus
we expect a total of one-tenth of a percent of all radio sources
to show this sort of extreme bend.

Bent jets seem to be more common than this. It would appear that the
conditions to produce bends are more common than we have assumed.  The
jet may interact with more than one cloud, extending the lifetime and
the likelihood of observation.  More detailed simulations and better
statistics of such bends may allow us to estimate the number of
sources with a sufficiently clumpy medium to make collisions likely.
An observational study of the true statistics of bent jets may allow
us to predict the number of sources with a sufficiently clumpy medium
to make collisions likely.

We have simulated the passage of a jet through a medium containing an
ensemble of clouds in \cite{HOD99,H98}. As the jet progresses
through the grid it is deflected several times where it has
encountered clouds, but can clearly be defined through the chain of
knots, whose total lifetime is much longer.

\section*{Acknowledgments}

SWH acknowledges the PPARC for receipt of a studentship. Computing was
performed using the Liverpool John Moores University Starlink node. We
 thank Prof.\ Sam Falle 
and Dr Alan Heavens for valuable suggestions.



\end{document}